\documentclass[twocolumn,a4paper]{revtex4}
\usepackage{comment}
\usepackage{color}
\usepackage{amsmath}
\usepackage{amssymb}
\usepackage{graphicx}
\usepackage{braket} 
\usepackage{esint}
\usepackage{adjustbox} 
\usepackage[utf8]{inputenc}

\usepackage[unicode=true,pdfusetitle,
bookmarks=true,bookmarksnumbered=false,bookmarksopen=false,
breaklinks=true,pdfborder={0 0 1},backref=false,colorlinks=true]
{hyperref}
\hypersetup{
	linkcolor=red,urlcolor=blue,citecolor=blue,anchorcolor=blue} 
\usepackage{ulem} 

\usepackage{dcolumn}\usepackage{bm}

\makeatother

\makeatother

\begin{document}

\title{Gray/dark soliton behavior and population under a symmetric and asymmetric potential trap}

\author{Jameel Hussain}

\affiliation{Department of Electronics, Quaid-i-Azam University, Islamabad, Pakistan}

\author{Javed Akram}

\email{javedakram@daad-alumni.de}

\affiliation{Department of Physics, COMSATS Institute of Information Technology, Islamabad, Pakistan}

\author{Farhan Saif}
\affiliation{Department of Electronics, Quaid-i-Azam University, Islamabad, Pakistan}

\date{\today}
\begin{abstract}
We numerically study the impact of Gaussian barrier height and width on gray solitons population in a symmetric and asymmetric potential trap. 
The gray solitons are created in a double-well potential by density engineering method.
Two identical Bose-Einstein condensates fragments are confined and made to collide by switching off the Gaussian barrier in a double-well potential. 
We find that the gray solitons population can be manipulated by Gaussian barrier height and width. We also study the 
gray solitons population dependence on the coupling strength. Moreover, we also study the impact of an asymmetry present in the
double-well potential. We observe that such an asymmetry always swings the point of collision of the gray solitons. Later, a stationary
dark soliton is created by phase imprinting method and we observe that the initial asymmetry in the double-well potential trap sets the dark soliton into oscillation.
\end{abstract} 

		
\maketitle

\section{Introduction} 
Soliton, a particle like wave behavior that shows a shape stability due to a balance between dispersion and non-linearity, formed in a vast 
range of physical systems from water-channels \cite{russell1845report}, optical fiber \cite{PhysRevE.93.032221,PhysRevE.89.011002} to Bose-Einstein condensates 
(BECs) \cite{Strecker02,Khaykovich1290,PhysRevLett.96.170401,PhysRevLett.83.5198,Denschlag2000,PhysRevLett.86.2926,1555-6611-26-6-065501,PhysRevA.93.033610,1612-202X-15-2-025501}. 
These are observed on a wide range of scale, such as on our wrist as a pulse \cite{1987506} and as a fast moving tsunami in the oceans \cite{Constantin-2009,Hereman2011}. 
Scott Russell in $1834$ observed this phenomenon in a water-channel and named  as ``\textit{wave of translation}'' 
\cite{russell1845report}. The Korteweg and de Vries (KdV) equation \cite{darrigol2005worlds} explained the theoretical 
nature of soliton while Zabusky and Kruskal gave the numerical proof that solitons also preserve their shape in collisions \cite{PhysRevLett.15.240,PhysRev.168.124}. 
\par
The Bose-Einstein condensate proposed theoretically \cite{clark1984einstein} and 
realized experimentally \cite{Anderson-1995,PhysRevLett.75.3969}.
The experimental realization of BEC in 1995 opened up a door to observe the quantum effect on macroscopic scale. 
Solitons are the excitation on the non-linear system and hence the BECs being a pure non-linear and a well controllable system is a 
fascinating  ground to study the soliton behavior. Non-linear Schr\"{o}dinger equation (NLSE), also known as Gross-Pitaevskii equation (GPE), 
describes the mean-field dynamics of the BECs \cite{RevModPhys.71.463}.
\par 
Solitons are classified as bright \cite{PhysRevLett.86.2926}, gray and black \cite{1751-8121-43-21-213001}. Bright soliton, a non-spreading matter-wave packet represents the 
ground state of the system  and has been observed in BECs with attractive interaction strength 
\cite{Strecker02,Khaykovich-2002,PhysRevLett.96.170401}. Dark/gray solitons are the excited states with energies greater than the ground 
state of the underlying system. These are  observed as a notch in the density distribution and has been observed in BEC with repulsive 
interaction strength \cite{PhysRevLett.83.5198,Denschlag2000,PhysRevLett.86.2926}. 
Solitons can be created by matter-wave density engineering 
method in a double-well potential \cite{PhysRevLett.95.010402,PhysRevA.73.013604,PhysRevA.78.063604}. The double-well potential trap is  
created by adding a Gaussian barrier at the center of the harmonic potential. The height $V_0$ and width $\sigma$ of the Gaussian barrier 
are defined as system parameters. These parameters controls the separation between the two fragments of the BECs.
\par
In this paper, we numerically investigate the impact of the barrier height $V_0$ and width 
$\sigma$ on the gray solitons population. We also aim to study the impact of  asymmetry in the
double-well potential on the dark solitons. In addition, we observe the creation of dark soliton by 
phase imprinting and density engineering methods. A phase difference of  exactly $\pi$ creates a dark soliton or $\pi$ state \cite{Denschlag2000,PhysRevLett.83.5198,PhysRevA.60.R3381}.
\par
This paper is organized as follows. We present the theoretical model in section II. Later, we describe the numerical 
 procedure and discuss our numerical results in section III. In section IV, we discuss the summary of our numerical results and our conclusions.

\section{Theoretical Model}
For a theoretical purpose, in the mean-field regime, the quasi one-dimension (1D) GPE models BEC near absolute zero temperature limit  
 \cite{Gross-1963,citeulike:7001053} 
\begin{eqnarray}
\iota\hbar\frac{\partial\psi(x,t)}{\partial t}=\left[-\frac{\hbar^{2}}{2m}\frac{\partial^{2}}{\partial x^{2}}+V_{ext}+g_{s}|\psi(x,t)|^{2}\right]\psi(x,t),
 \label{eq:1}
\end{eqnarray}

where $\psi(x,t)$ represents wave function of the BEC with normalization condition $\int|\psi(x,t)|^{2}dx=1$, $m$ represents mass of individual atom, 
$t$ and $x$ stands for time and 1D-space coordinate respectively. The BECs experience the external potential $V_{ext}=V_h+V_b+V_a$, with $V_h$ represents 
harmonic potential part, $V_b$ describes the Gaussian barrier potential part, which converts simple harmonic trap into double-well potential trap. 
Such potential help to separate and confine the BECs into two identical fragments. While $V_a$ stands for the asymmetry in the double-well potential. The interaction 
strength $g_{s}=2N\hbar\omega_{r}a_{s}$ characterizes by s-wave scattering length 
$a_{s}$ \cite{PhysRevA.93.033610}, $N$ represents the total number of atoms in
BECs and $\omega_{r}$ defines the radial frequency component of the harmonic trap \cite{PhysRevA.93.023606}.

\par
The quasi-1D GP equation $(1)$ reduces to the dimensionless form by measuring time, 
length and energy in units of $\omega_{x}^{-1}$, $\sqrt{\hbar/m(\omega_{x})}$ and $\hbar\omega_{x}$ respectively as,

\begin{eqnarray}
\iota\frac{\partial\psi}{\partial t}=\left[-\frac{1}{2}\frac{\partial^{2}}{\partial x^{2}}+V_{ext}+g_{s}|\psi|^{2}\right]\psi 
\label{eq:2}
\end{eqnarray}
where, the dimensionless 1D interaction strength is described as $g_{s}=2N\omega_{r}a_{s}/(\omega_{x}L)$, here $L$ represents the length and the dimensionless external potential is given by 
$V_{ext}=x^{2}/2+V_{0}.e^{-(x/\sigma)^2}+V_{1}.x.U(x)$, with $U(x)$ is defined as a dimensionless unit step function.

\begin{figure}
\includegraphics[height=10cm,width=9cm]{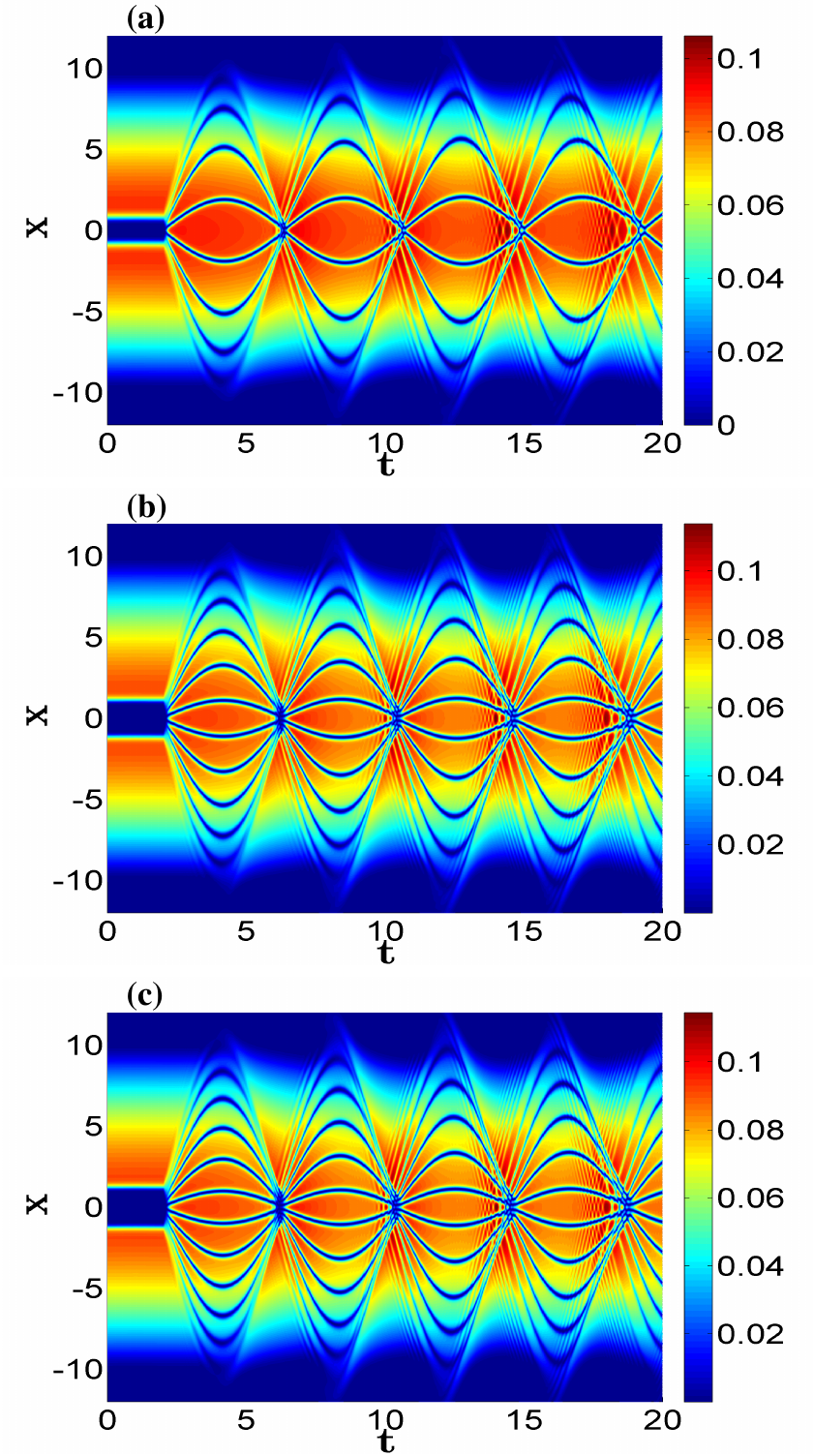} 
\label{Fig:1}
\caption{(Color online) Gray solitons population growth vs the barrier height $V_0$. The gray solitons are created 
by density engineering method.
The dimensionless parameter are  coupling strength $g_s=500$, Gaussian barrier height (a) $V_0=100$, (b) $V_0=500$, (c) $V_0=900$ and for a constant barrier width 
$\sigma=0.632$. Even number of gray solitons  $8$, $10$ and $12$ are created for $V_0=100$, $V_0=500$ 
and $V_0=900$ respectively from  top to bottom.}
\end{figure}

\section{Numerical Procedure}
We create gray solitons numerically by density engineering method, in a collision between two identical BEC fragments.
Both fragments are separated and confined by a double-well potential trap. Such a double-well potential trap is created by adding a
Gaussian barrier  $V_b=V_0e^{-({\frac{x}{\sigma}})^2}$ at the center of a harmonic potential $V_h=x^2/2$. 

\par
For numerical simulations we do discretization of the dimensionless GP equation (2), 
with space step $\Delta{x}=0.0177$ and the time step $\Delta{t}=0.0001$. The time-splitting spectral method 
\cite{BAO2003318,Vudragovic12,Kumar15,Loncar15,Sataric16} is used to find the ground state wave function of the BECs  by simulating in imaginary 
time $\tau=\iota t$. Initially, we drive test wave function $\psi(x)=\sqrt{(\mu-V_{ext})/g_s}$ after Thomas-Fermi 
Approximation \cite{1555-6611-26-6-065501}, where $\mu$ represents the dimensionless chemical potential. For 
dynamical evolution of the wave function of the BECs fragments the ground state wave function served as an initial 
condition for rest of the  numerical simulations. 
 
\subsection{Impact of barrier height and width on gray solitons population}

\subsubsection{Symmetric double-well potential}
Two identical BECs fragments are prepared and trapped into a symmetric double-well potential. The trap potential is  
given by $V_{ext}(x)=x^{2}/2+V_{0}e^{-\frac{x^{2}}{\sigma}}$. Both the fragments are made to collide by quenching i.e, by switching off the Gaussian barrier $V_b$. 
After collision the BECs fragments are evolved in real time for a leftover harmonic 
potential. The collision is resulted in the creation of the density notches of different depths, known as gray solitons. 
The separation between the two BECs fragments is controlled by the barrier height $V_0$ and the barrier width $\sigma$. 
As a first case, we study the impact of the barrier height on the gray soliton's population as shown in Fig.~1. Moreover, we see that the soliton population shows its 
dependence on the initial barrier height $V_0$ as shown in Fig.~1. We observe $8$, $10$ and $12$ gray solitons for a barrier height of $100$, $500$ and $700$ respectively in Fig.~1, 
thus gray soliton's population increases with the barrier height. All the solitons start oscillations in such a way that they always reach the point of collision (POC) 
simultaneously. While on collision, they preserve their shape as shown in Fig.~1. Such kind of shape preserving collision is an intrinsic property of a 
matter wave soliton \cite{PhysRevLett.15.240,PhysRev.168.124,Frantzeskakis10}. In Fig.~1, we observe that the change in barrier height does not change the POC of gray solitons. 
This happens due to the initial trap symmetry.
\par
We also observe that for a fixed value of interaction strength $g_s=100$, the gray soliton's population increases rapidly and then get saturated 
for specific value of $\sigma$ as shown in Fig.~2. Furthermore, for a fixed value of Gaussian barrier height $V_0$ and barrier width $\sigma$, the gray soliton's population 
shows a direct dependence on the interaction strength $g_s$. However, for $g_s=500$, and $700$, the saturation point is out of the range of the graph given in Fig.~2. The saturation 
in the soliton's population results due to the finite mass (negative) of each gray soliton. Since for a constant interaction strength $g_s$, BEC has finite mass or number of 
constituent particles so it can sustain only finite number of solitons. 
\begin{figure}
\includegraphics[height=6.5cm,width=8.5cm]{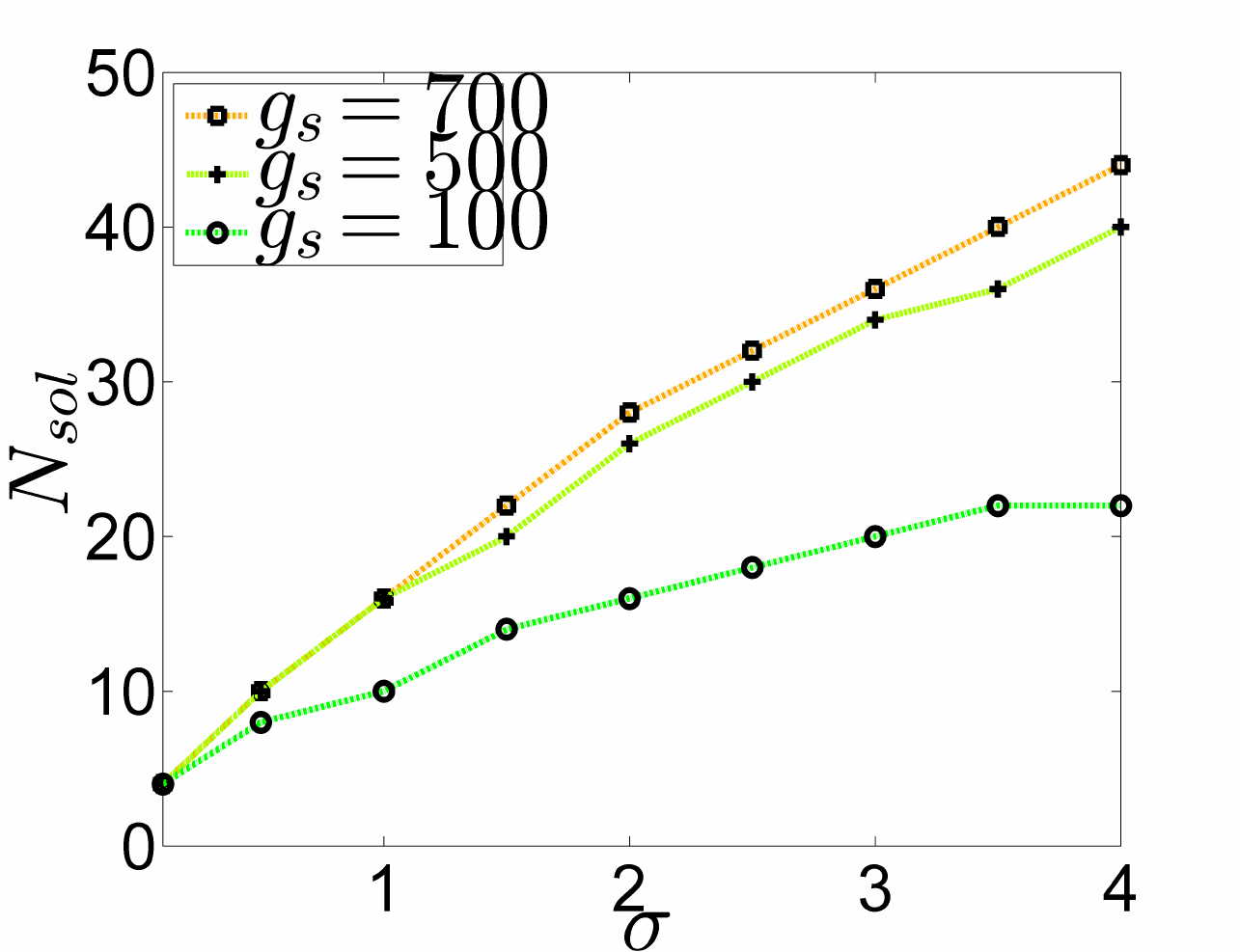}
\label{Fig:2}
\caption{(Color online) Gray soliton's population, $N_{sol}$ vs the barrier width $\sigma$. For a fixed value of the interaction strength, the gray soliton's population grows with the barrier width. Other dimensionless parameters are barrier height $V_0=500$, 
and interaction strengths in descending order are $g_s=700$, $g_s=500$ and  $g_s=100$.}
\end{figure} 

\subsubsection{Asymmetric double-well potential}

To observe the impact of asymmetry in the double-well potential on the gray soliton's population, we add an asymmetry $V_a$ in the double-well potential. Later, we numerically 
switch off both the Gaussian barrier potential and the asymmetry part of the potential. The system experiences a leftover harmonic potential. We observe that the presence of 
a small asymmetry in the double-well potential sets the POC of gray soliton into oscillations as depicted in Fig.~3. Hence an asymmetry introduces another 
oscillations in the system. We also observe that a small asymmetry introduced in double-well potential has no significant 
impact on the soliton's population as we can see in Fig.~3(a-c). The oscillations in 
POC is a direct consequence of an initial asymmetry present in the trapping potential. The amplitude of the oscillations for POC depends upon the amount of the asymmetry added, as we can see in Fig.~3.
\par 
\begin{figure}
\includegraphics[height=10cm,width=9cm]{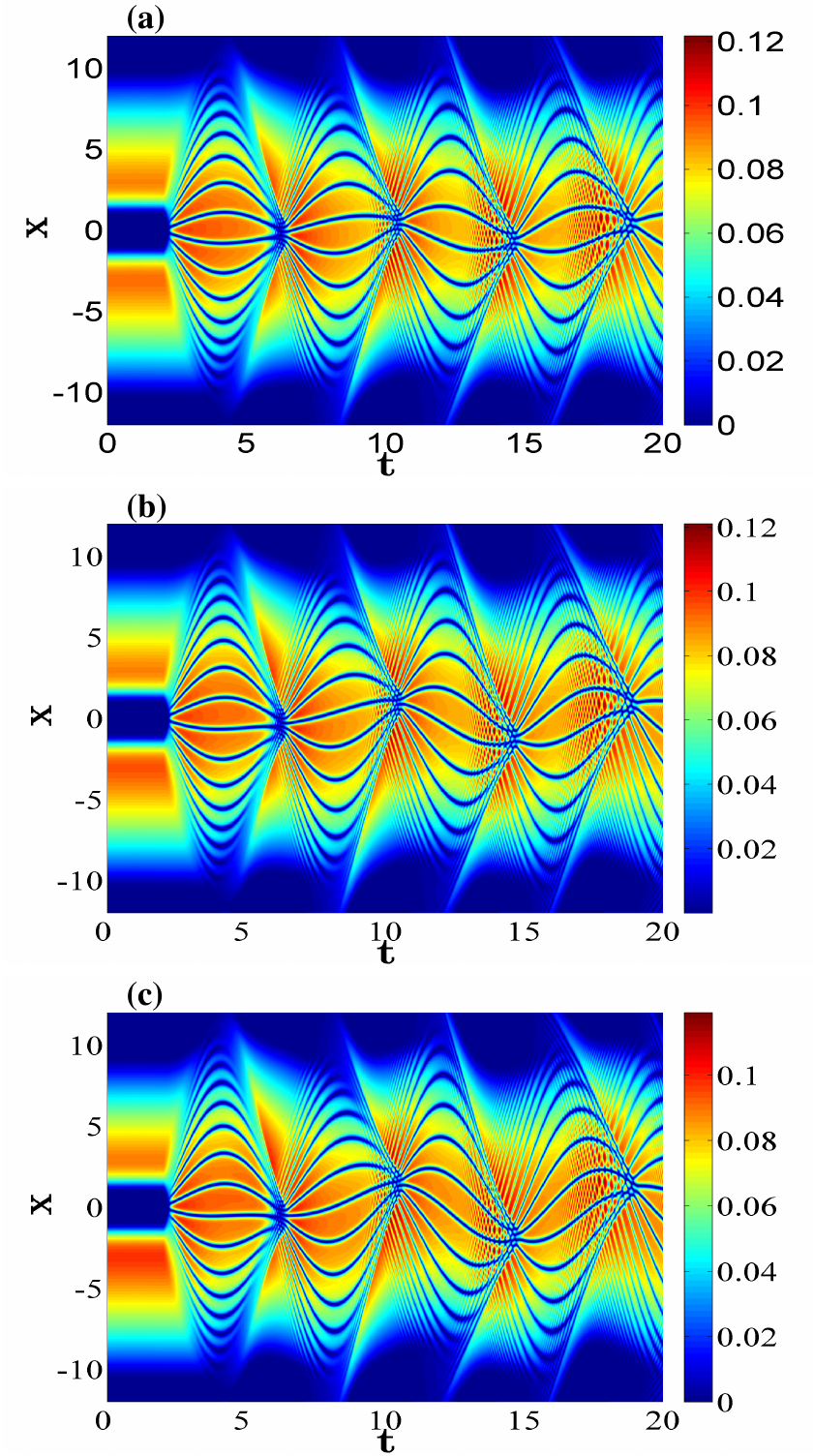} 
\label{Fig:3}
\caption{(Color online) The point of collision of gray solitons set into oscillation by adding a small 
asymmetry $V_a$ in the double-well potential trap. Here the dimensionless 
parameters are interaction strength $g_s=500$, barrier height $V_0=100$, barrier width $\sigma=0.632$ and asymmetry (a) $V_1=0.5$ (b) $V_1=1$ and (c) $V_1=1.5$. While 
asymmetry shows no impact on the gray solitons population. The amplitude of oscillation of POC proportional to the amount of asymmetry added.}
\end{figure}
\par 

\subsection{Oscillations in the dark soliton}
In this section, we demonstrate that for a phase difference of $\pi$ between two colliding BECs fragments a dark soliton 
is created at the center of the harmonic trap as shown in Fig.~4. To get system in equilibrium, initially we prepare our system in 
symmetrical double-well potential with the Gaussian potential barrier height $V_0=70$ and the barrier width $\sigma=0.632$. 
To create a dark soliton, we imprint a phase $\pi$ on one of the two BECs fragments in real time dynamics for a 
dimensionless time $t=2$ and then we switched off the Gaussian barrier. In Fig.~4(a), the central straight 
line shows a dark soliton and it has zero velocity. The dark soliton is supposed to be still due 
to its heavy negative mass \cite{Frantzeskakis10}. For a phase difference of zero, solitons are 
even in number as shown in Fig.~1 and for a phase difference of $\pi$ the number of gray/dark solitons are odd in number as depicted in Fig.~4.
 
\par
In Fig.~4(a) the central line represents a stationary dark soliton and in Fig.~4(b) 
that dark soliton sets into motion (oscillation). This happens when an asymmetry 
$V_1=2$ in the trapping potential is added for one of the two fragments along with a phase $\pi$ imprinted numerically on that fragment. We observe that the dark
soliton sets into oscillations. Oscillation in the dark soliton shows that the dark soliton is turned into a gray soliton. Here once again the 
POC starts swinging due to the presence of an asymmetry in the initial double-well potential. As we observed in the previous section  
that the presence of an asymmetry in the trapping potential always puts the POC into oscillation. 

\begin{figure}
\includegraphics[height=7cm,width=9cm]{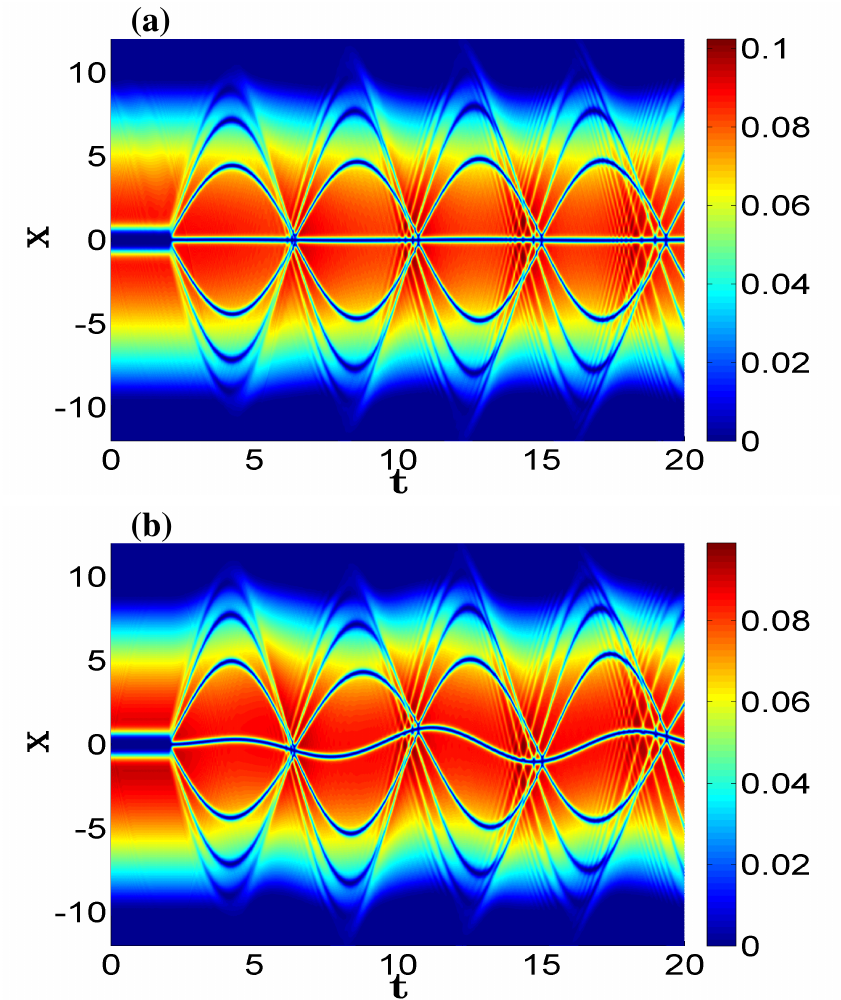} 
\label{Fig:4}
\caption{(Color online) (a) A stationary dark soliton, straight line at the center of the top figure, created by imprinting a 
phase $\pi$ on one of the two BEC fragments confined in a symmetric double-well potential trap. (b) The oscillation appears in the dark soliton by 
introducing a small dimensionless asymmetry $V_1=1$ in the double-well potential confining the two fragments. 
Other dimensionless parameters are interaction strength $g_s=500$, barrier height $V_0=70$ and barrier width $\sigma=0.632$. Here, 
odd number of soliton are observed in the presence of dark soliton.}
\end{figure}

\section{conclusion}
In this paper, we conclude that for a constant interaction strength $g_s$ the gray soliton's population can be manipulated by the Gaussian 
barrier height $V_0$ and/or by Gaussian barrier width $\sigma$. Gray soliton's population increases with the increase in 
barrier height and barrier width. Additionally, we observe a direct dependence of soliton's population on interaction 
strength for a constant height and width of the Gaussian potential barrier. 
\par
It is known that dark soliton has zero velocity due to its heavy negative mass. We observe that a dark 
soliton can be set into oscillations by introducing an asymmetry in the  
confining double-well potential. The dark soliton can be converted into gray one by adding an asymmetry in the 
double-well potential. We observe that addition of a small asymmetry always set the point of collision (POC) 
into oscillations without influencing the gray solitons population. The amplitude of oscillations of POC is proportional to the amount of asymmetry in the double-well potential. 
Hence the oscillation in the POC can be used as tool to quantify the presence of an initial asymmetry in the double-well potential trap.
\par 

\section{Acknowledgment}
Jameel Hussain gratefully acknowledges support from the COMSATS University Islamabad for providing him a workspace. 
 
\bibliographystyle{apsrev4-1}


%

\end{document}